\documentclass[pdflatex,sn-basic]{sn-jnl}

\jyear{2021}%

\theoremstyle{thmstyleone}%
\theoremstyle{thmstyletwo}%

\theoremstyle{thmstylethree}%

\renewcommand{\vec}[1]{\mbox{\boldmath $#1$}}

\newcommand*{\doi}[1]{\href{http://dx.doi.org/#1}{doi: #1}}

\def\pa{\partial}
\def\vb{{\bf v}}
\def\Vb{{\bf V}}

\def\Bb{{\bf B}}
\def\vf{{\bf V}_m}
\def\Bf{{\bf B}}

\def\ep{{\bf e}_{\phi}}

\def\MC{meridional circulation}
\begin{document}

\title[Mean field models]{Mean field models of flux transport dynamo and
meridional circulation in the Sun and stars}


\author*[1,2]{\fnm{Gopal} \sur{Hazra}}\email{hazra@iitk.ac.in}

\author[3,4]{\fnm{Dibyendu} \sur{Nandy}}\email{dnandi@iiserkol.ac.in}
\equalcont{These authors contributed equally to this work.}

\author[5]{\fnm{Leonid} \sur{Kitchatinov}}\email{kit@iszf.irk.ru}
\equalcont{These authors contributed equally to this work.}

\author[6]{\fnm{Arnab} \sur{Rai Choudhuri}}\email{arnab@iisc.ac.in}
\equalcont{These authors contributed equally to this work.}

\affil*[1]{\orgdiv{Dept. of Physics}, \orgname{Indian Institute of Technology, Kanpur}, \orgaddress{\street{Kalyanpur}, \city{Kanpur}, \postcode{208060}, \state{Uttar Pradesh}, \country{India}}}

\affil[2]{\orgdiv{Dept. of Astrophysics}, \orgname{University of Vienna}, \orgaddress{\street{T\''urkenschanzstraße 17}, \city{Vienna}, \postcode{1180}, \state{Vienna}, \country{Austria}}}

\affil[3]{\orgdiv{Department of Physical Sciences}, \orgname{Indian Institute of Science Education and Research, Kolkata}, \orgaddress{\street{Mohanpur}, \city{Kolkata}, \postcode{741 246}, \state{WB}, \country{India}}}

\affil[4]{\orgdiv{Center of Excellence in Space Sciences India}, \orgname{Indian Institute of Science Education and Research, Kolkata}, \orgaddress{\street{Mohanpur}, \city{Kolkata}, \postcode{741 246}, \state{WB}, \country{India}}}

\affil[5]{\orgdiv{Institute of Solar-Terrestrial Physics} \orgname{SB RAS}, \orgaddress{\street{Lermontov Str. 126a}, \city{Irkutsk}, \postcode{664033}, \state{}\country{Russia}}}

\affil[6]{\orgdiv{Dept. of Physics}, \orgname{Indian Institute of Science}, \orgaddress{\street{C V Raman Avenue}, \city{Bengaluru}, \postcode{560012}, \state{Karnataka}, \country{India}}}


\abstract{The most widely accepted model of the solar cycle is the flux transport dynamo model. This model evolved out of the traditional $\alpha \Omega$ dynamo model which was first developed at a time when
the existence of the Sun's meridional circulation was not known. In these models the toroidal magnetic field (which gives rise to sunspots) is generated by the stretching of the poloidal field by solar differential rotation. The primary source of the poloidal field in the flux transport models is attributed to the Babcock--Leighton mechanism, in contrast to the mean-field $\alpha$-effect used in earlier models. With the realization that the Sun  has a meridional circulation, which is poleward at the surface and is expected to be equatorward at the bottom of the convection zone, its importance for transporting
the magnetic fields in the dynamo process was recognized. Much of our understanding about the physics of both the meridional circulation and the flux transport dynamo has come from the mean field theory obtained by averaging the equations of MHD over turbulent fluctuations. The mean field theory of meridional circulation makes clear how it arises out of an interplay between the centrifugal and thermal wind terms.  We provide a broad review of mean field theories for solar magnetic fields and flows, the flux transport dynamo modelling paradigm and highlight some of their applications to solar and stellar magnetic cycles. We also discuss how the dynamo-generated magnetic field acts on the meridional circulation of the Sun and
how the fluctuations in the meridional circulation, in turn, affect the solar dynamo. We conclude with some remarks on how the synergy of mean field theories, flux transport dynamo models and direct numerical simulations can inspire the future of this field.}




\maketitle

\section{Introduction}\label{sec1}

The turbulent convection zones of the Sun
and other stars host a magnetohydrodynamic dynamo mechanism which involve interactions between the velocity and the magnetic fields. When this interaction takes place within a rotating astrophysical object,
it leads to the possibility of a large-scale magnetic field emerging out of such interactions\citep{Parker55a,SKR66,Moffatt_78,Parker_79}. Historically this subject developed by solving the mean
field equations which arise by averaging over turbulence at small scales. The challenge of the subject comes from the fact that physics at small scales may profoundly influence what is happening at the large scales. The
physics of small scales is captured in the mean field equations through a set of parameters—the $\alpha$-effect, turbulent diffusion, Reynolds stresses (including what is called the $\Lambda$-effect), and turbulent pumping. We shall collectively refer to them
as ‘turbulence parameters’ \citep{Moffatt_78}. The mean field theory has two aspects.
(i) We have to estimate the turbulence parameters by some means.
(ii) We have to solve the mean field equations in which these turbulence parameters appear.
In the early years of research, turbulence parameters would be calculated analytically by making some suitable assumptions about the small-scale turbulence \citep[e.g.,][]{Chou98}. Sometimes, observational data could be used to put important constraints on
these parameters \citep[e.g.,][]{Chae2008,HM18}. Within the last few years, it has been possible to calculate
the turbulence parameters from numerical simulations of turbulence \citep[e.g.,][]{Kapyla06,Simard_Charbonneau_2016,Shimada_Hotta_2022}. We expect
more inputs from simulations in the coming years to put the mean field models
on a firmer footing.
There is a common consensus that the mean field models played a historically important role in the development of the subject.  However, with increasingly complex and computationally intensive full magnetohydrodynamic (MHD)
simulations being done by various groups around the world, are mean field models still relevant?

The mean field modelling approach is computationally less demanding and it is possible to make more extensive parameter space studies with them.
However, there is a deeper reason why the mean field
models continue to remain so relevant. Even
the most ambitious numerical simulations undertaken at the present time have
fluid and magnetic Reynolds numbers many orders of magnitude smaller than
what they are for the Sun and other stars. They are still very far from producing sufficiently realistic results which can be compared with observational data
in detail.  On the other
hand, by adjusting various parameters of mean field models suitably, it is often
possible to achieve remarkable agreements with observations. This approach
may justifiably be criticized as ad hoc. However, an understanding of what
needs to be done in the mean field models to achieve convergence with observational data often provides great insights into various physical processes. Mean
field models are expected to remain an active research area for many years to
come.

The mean field theory of large-scale magnetic fields is easily adapted to an approach known as kinematic dynamo theory \citep{Wang91,CSD95}. For solving the equations of kinematic dynamo theory, we have
to specify the large-scale flows—such as the differential rotation and the meridional circulation—apart from the turbulence parameters. While the kinematic
dynamo theory was developing, important developments also took place in the
mean-field theory of large-scale flows—the initial impetus coming from efforts
to explain the differential rotation of the Sun. For a few decades, the kinematic
dynamo theory and the mean field theory of large-scale flows developed almost
independently of each other. These two theories have come together in the
last few years with the blossoming of the field of solar and stellar dynamos. Kinematic
models of the solar dynamo could rely on the observations of the differential
rotation and the meridional circulation of the Sun. We do not have similar
detailed observational data of other stars. In order to build mean field models
of stellar dynamos, the kinematic dynamo theory and the mean field theory of
large-scale flows have to be combined together. In the case of the Sun also, as we
have become aware of various feedback processes between the solar magnetic
cycle and the large-scale flows, it has become essential to combine the two theories of the kinematic dynamo and large-scale flows—to model such phenomena as
torsional oscillations and variations of the meridional circulation.

The first models of the solar dynamo were constructed at a time when the
only available knowledge about large-scale flows was the existence of differential
rotation at the solar surface. Nothing was known about the meridional circulation of the Sun or the distribution of differential rotation underneath the solar
surface. With the discovery of the meridional circulation and the realization that
it is likely to play an important role in the dynamo process, a new type of dynamo model—the flux transport dynamo model—came into being. There are
now efforts of applying the flux transport dynamo model to other stars. Since
the meridional circulation plays a crucial role in the flux transport dynamo, the
temporal variation of this circulation in the Sun may have a profound effect on
the solar cycle. Some of the irregularities of the solar cycle seem to arise from
fluctuations in the meridional circulation.

The solar dynamo has been the subject of several reviews \citep{Chou11,Char14,Karak14,Charbonneau2020}.  The prospect for extrapolating the
solar dynamo models to stars has been reviewed by \citet{Chou17}. We also
refer to a recent review of the meridional circulation \citep{Chou21b}.

The mean field models of large-scale magnetic and velocity fields are described in the next section. Then section~\ref{sec3} describes how large-scale flows in the Sun
and stars—especially the meridional circulation—can be computed from the
mean field model of large-scale flows. How the flux transport dynamo model
arose for explaining the solar cycle will be discussed in the section~\ref{sec4}, with section~\ref{sec5} summarizing
its applications to other stars. Then section~\ref{sec6} will point out the key role of meridional
circulation variations in explaining the solar cycle irregularitis. In the section~\ref{sec7}, we survey
the current status of the important subject of computing the turbulent parameters from numerical simulations, which may provide important inputs to mean
field models, and conclude our review.

\section{Mean-field theory of large-scale magnetic and velocity fields}\label{sec2}

Magnetic and velocity fields of the Sun behave differently on large and small spatial scales. The fields of the scale comparable to the solar radius show repeatable---though not strictly
periodic---evolution of their patterns in the course of 11-year solar cycles \citep{Hathaway_15}. Cells of granular or supergranular solar convection, on the other hand, reconfigure themselves
irregularly on a much shorter time scale. This leads to the basic idea of mean-field magnetohydrodynamics (MHD) that not detailed structures but mean statistical properties only of the  small-scale turbulent magnetic ($\vec{b}$) and velocity ($\vb$) fields are relevant to
the dynamics of their large-scale magnetic ($\vec{B}$) and velocity ($\vec{V}$) counterparts. Different scales are separated by temporal or spatial averaging, which leaves the large-scale field unchanged but nullifies the turbulent fields: $\langle\vec{b}\rangle = 0,\ \langle\vb \rangle = 0$, where the angular brackets signify the averaging.

Formulation of the mean-field MHD is a formidable task that is not completed up to now. Nevertheless, the main effects and methods of the mean-field theory were systematised already about forty years ago in the monographs by \citet{Moffatt_78}, \citet{Parker_79} and \citet{Krause_Raedler_80}. The mean field induction equation of the theory,
\begin{equation}
    \frac{\partial{\vec B}}{\partial t} = {\vec\nabla}\times
    \left({\vec V}\times{\vec B} + {\vec{\cal E}}\right),
    \label{induction}
\end{equation}
includes the small-scale turbulent fields via the Mean Electromotive Force (EMF) ${\vec{\cal E}} = \langle \vb \times{\vec b}\rangle$. In what follows, we use the following basic expression for the EMF
\begin{equation}
    {\vec{\cal E}} = \alpha{\vec B} + {\vec v}^{\rm dia}\times{\vec B}
    - \eta_{_{\rm T}}{\vec\nabla}\times{\vec B} ,
    \label{EMF}
\end{equation}
which includes so-called $\alpha$-effect, diamagnetic pumping and eddy diffusion, respectively, in its right-hand side.

Among the three effects, diamagnetic pumping is the least known one. The diamagnetic effect of turbulent conducting fluids consists in expulsion of magnetic fields from the regions of relatively high turbulent intensity \citep[see][for pictorial explanation]{Kit_Ole_12Dyn}. The effect was predicted by \citet{Zeldovich_Dia_57} and first derived by \citet{Radler_Dia_68} where the expression for the effective velocity
\begin{equation}
    {\vec v}^{\rm dia} = -\frac{1}{2}{\vec\nabla}\eta_{_{\rm T}}
    \label{V_dia}
\end{equation}
can be found \citep[see also Eq.\,(3.10) in][]{Kit_Rue_92Dia}. The diamagnetic pumping has been detected in MHD laboratory experiment \citep{Spence_EA_07Dia} and in direct numerical simulations \citep{Tobias_EA_98Dia_by_DNS, Dorch_Nordlund_01Dia,Ossendrijver_EA_02Dia}. If included in a dynamo model, the downward diamagnetic pumping with the effective velocity of Eq.\,(\ref{V_dia}) can concentrate magnetic fields to the bottom of the convection zone thus realising an interface dynamo even in distributed-type models  \citep{Kit_Ole_12Dyn}.

The coefficient $\alpha$ appearing in Eq.\,(\ref{EMF}),
which becomes a rank-2 tensor in the
completely general situation, arises from helical turbulent motions and is crucial for the
dynamo generation of the magnetic field. It was first evaluated by \citet{Parker55a} and
\citet{SKR66}. For isotropic turbulence, suitable assumptions lead to the
expression
\begin{equation}
    \alpha = - \tau \langle \vb \cdot ({\vec\nabla} \times \vb ) \rangle /3,
    \label{alpha}
\end{equation}
where $\tau$ is the correlation time of turbulence \citep{Chou98}. As we shall point
out in section~\ref{sec4}, the flux transport dynamo involves a different mechanism for
magnetic field generation: the Babcock--Leighton mechanism.  This mechanism also can
be represented by the coefficient $\alpha$ appearing in Eq.\,(\ref{EMF}). However,
$\alpha$ corresponding to this mechanism is not given by Eq.\,(\ref{alpha}).

It has been found relatively recently that the conservation of magnetic helicity leads to
a catastrophic quenching of the $\alpha$-effect \citep{Gruzinov_Diamond_94,Brandenburg_Subramanian05}, switching off the local $\alpha$-effect of Eq.\,(\ref{EMF}) in the case of large magnetic Reynolds number. It can be shown that non-local $\alpha$-effect of Babcock-Leighton type is not subject to the catastrophic quenching \citep{Kit_Ole_11Alleviation}.

We now turn to the mean field theory of large-scale flows. The mean-field  equation of motion is
\begin{equation}
    \frac{\partial{\vec V}}{\partial t}
    + ({\vec V}\cdot{\vec\nabla}){\vec V} =
    \frac{1}{\mu\rho}({\vec B}\cdot{\vec\nabla}){\vec B}
    -\frac{1}{\rho}{\vec\nabla}\left(P + \frac{B^2}{2\mu}\right)
    + {\vec g} + \frac{1}{\rho}{\vec\nabla}\cdot{\vec{\cal S}},
    \label{motion}
\end{equation}
where $P$ is pressure, $\vec g$ is gravity and
\begin{equation}
    {\cal S}_{ij} = -\rho\langle {\rm v}_i{\rm v}_j\rangle
    + \mu^{-1}\langle b_ib_j - \frac{1}{2}\delta_{ij}b^2\rangle
    \label{stress}
\end{equation}
is the turbulence stress tensor combining the Reynolds and Maxwell stress of small-scale fields.

Mean-field MHD has to express the turbulent stress in terms of large-scale fields. Galilean invariance demands the mean velocity to contribute via its spatial derivatives only:
\begin{equation}
    {\cal S}_{ij} = \rho N_{ijkl}\nabla_kV_l
    \label{stress1}
\end{equation}
(repetition of subscripts means summation). This linear relation can be seen as a formal definition of the eddy viscosity tensor $N_{ijkl}$. Enhanced dissipation of large-scale flows is a well-known effect of turbulence. This is however not all what convective turbulence in stars can do. Convection is driven by buoyancy forces which point upward or downward in radius. This imparts {\em anisotropy} with different intensities of radial and horizontal turbulent mixing. It is known since \citet{Lebedinsky_41} that the eddy viscosity tensor for anisotropic turbulence does not satisfy the Onsager symmetry rule $N_{ijkl} = N_{klij}$. The rule has to be satisfied for true viscosity decreasing kinetic energy of mean flow \citep[see Sect.\,I.9 in][] {Landafshitz_Kinetics_81}. Violation of the rule signals that the anisotropic turbulence does not necessarily dissipate but can excite some kind of large-scale flow. An important application of this excitation effect was found in the theory of stellar differential rotation where it is known as the $\Lambda$-effect \citep{Ruediger_89}. The stress tensor of Eq.\,(\ref{stress1}) for anisotropic turbulence does not vanish for rigid rotation,
\begin{equation}
    {\cal S}^\Lambda_{ij} = \rho N_{ijkl}\varepsilon_{klm}\Omega_m
    \label{Lambda}
\end{equation}
where $\vec\Omega$ is the angular velocity and $\varepsilon_{klm}$ is the fully antisymmetric tensor. The components ${\cal S}^\Lambda_{r\phi}$ and ${\cal S}^\Lambda_{\theta\phi}$ in spherical coordinates stand for the angular momentum fluxes by turbulence, which are the principal drivers of stellar differential rotation. Details of the $\Lambda$-effect theory can be found in \citet{Ruediger_89} and \citet{Rue_Kit_Holl_13}. Separation of the $\Lambda$-effect from true viscosity changes Eq.\,(\ref{stress1}) to
\begin{equation}
    {\cal S}_{ij} = {\cal S}^\Lambda_{ij} + \rho {\cal N}_{ijkl}\nabla_kV_l ,
    \label{stress2}
\end{equation}
where ${\cal N}_{ijkl}$ is the true viscosity tensor with positive definite coefficients and symmetry, ${\cal N}_{ijkl} = {\cal N}_{klij}$, ensuring dissipation of large-scale flows. Derivation of the viscosity tensor for rotating turbulence can be found in \citet{KPR_94Diffusion}.

It may be noted that several effects in excess of the $\Lambda$-effect and eddy viscosity of Eq.\,(\ref{stress2}) have been found in the extensive literature on turbulent stress. These include turbulent pressure, a slight modification of the large-scale Lorentz force \citep{Klee_Rog_94LFmodif,RKS_12LFmodification}, and the anisotropic kinetic alpha-effect \citep{Frisch_EA_87AKA}. Equation (\ref{stress2}) includes what matters for stellar applications only. A similar comment applies to the EMF of Eq.\,(\ref{EMF}). The equation displays its three basic contributions in the simplest form. Rotationally induced anisotropy complicates them \citep{Pipin_08EMF,KPR_94Diffusion} so that, e.g., the eddy magnetic diffusivities for the directions along and across the rotation axis differ. The rotational anisotropy however is of modifying rather than principal nature for stellar dynamo modelling.
\section{Meridional circulation in the Sun and stars}\label{sec3}
The global flow in the sun is known to vary little in course of the activity cycle. The flow is not magnetic by origin. We consider first hydrodynamics of the meridional flow and discuss its magnetic modification afterwards.
\subsection{Meridional flow origin and structure}\label{sec3.1}
Momentum density in the stellar convection zones is divergence-free, ${\vec\nabla}\cdot(\rho{\vec u}) = 0$, to a good approximation \citep{Lanz_Fan_99Anelastic}. In this case, the meridional flow is a fluid circulation over {\em closed} stream-lines.

The circulation proceeds in a turbulent convection zone where the eddy viscosity resists the flow. Some forces supporting the flow against the viscous decay should therefore be present. Only non-conservative forces can transmit energy to a circulatory flow. Motion equation (\ref{motion}) can be curled to filter-out  irrelevant conservative forces: see,
for exmaple, \citet{Chou21b}. This leads to a meridional flow description in terms of the azimuthal vorticity $\omega = ({\vec\nabla}\times{\vec V})_\phi$:
\begin{equation}
    \frac{\partial\omega}{\partial t}
    + r\sin\theta\,{\vec\nabla}\cdot\left({\vec V}\frac{\omega}{r\sin\theta}\right)
    + {\cal D}(\omega)
    = r\sin\theta\frac{\partial\Omega^2}{\partial z} -
    \frac{g}{r c_{\mathrm p}}\frac{\partial S}{\partial\theta} .
    \label{vorticity}
\end{equation}
In this equation, $z = r\cos\theta$ is the (signed) distance from the equatorial plane, $S$ is the specific entropy, and $\cal D$ accounts for the viscous dissipation of the meridional flow. The symbolic representation for the dissipation term is justified by complexity of its explicit formulation \citep{Kit_Ole_11DR}. The dissipation term acts to decrease the meridional flow energy.

Two terms in the right-hand side of Eq.\,(\ref{vorticity}) stand for two principal drivers of the meridional flow. The first term includes driving by the centrifugal force. The force is conservative for cylinder-shaped ($z$-independent) rotation. Accordingly, the first term in the right-hand side of (\ref{vorticity}) accounts for the non-conservative part of the centrifugal force. The second term involves the non-conservative buoyancy (baroclinic) force.

\begin{figure}
    \centering
    \includegraphics[width= 7 truecm]{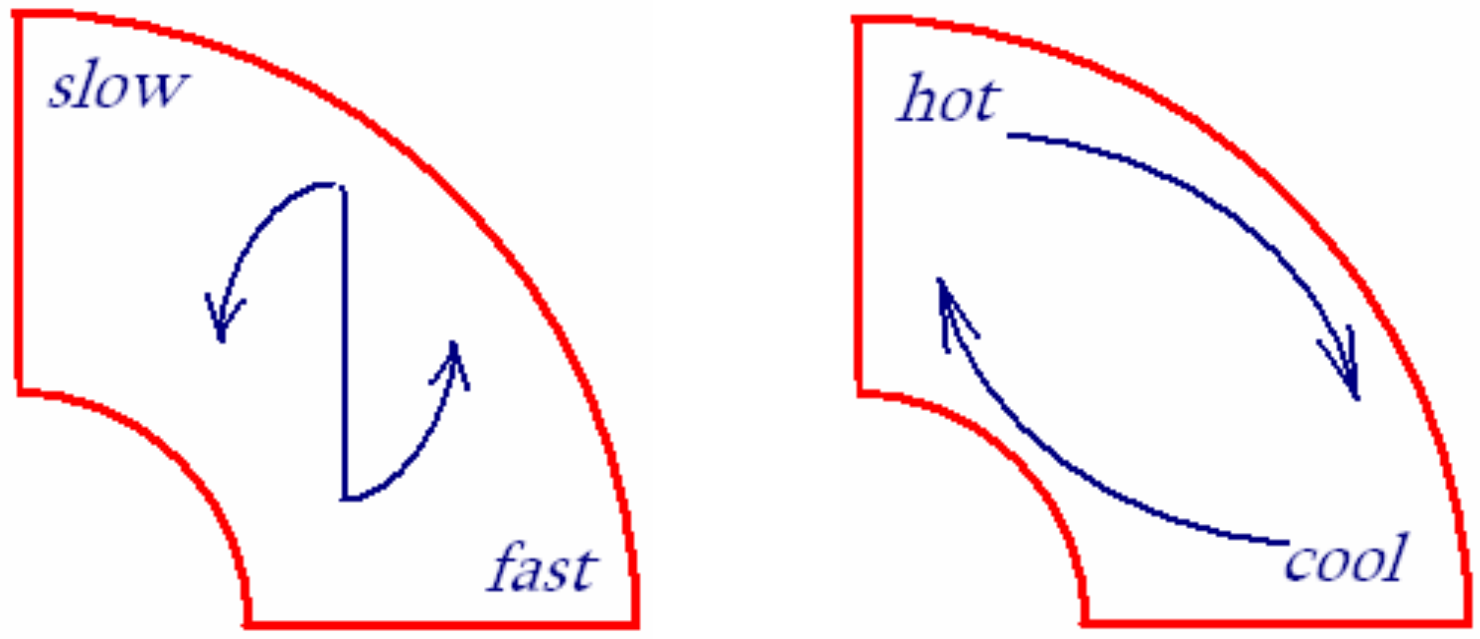}
    \caption{Illustration of the centrifugal (left) and baroclinic (right)
        driving of the meridional flow (see text).}
    \label{drivers}
\end{figure}

Figure\,\ref{drivers} illustrates the two drivers. If the angular velocity decreases with distance from the equatorial plane, as it does in the sun \citep{Schou_EA_98Helioseismic_DR}, a torque by the centrifugal force tends to drive anti-clockwise circulation (in the north-west quadrant of the convection zone). The baroclinic driving is proportional to the temperature variation with latitude inside the convection zone. The \lq differential temperature' results from rotationally-induced anisotropy of the convective heat transport \citep{Ruediger_EA_05Heat_flux}. If the temperature increases with latitude, as it probably does in the Sun \citep{Miesch_EA_06Diff_temperature,Kit_Ole_11DR}, a slightly cooler fluid at low latitudes tends to sink down and the warmer polar fluid tends to rise up and spread over the surface to drive a clockwise circulation (Fig.\,\ref{drivers}).

The two drivers of the meridional flow counteract each other in the Sun. The counteraction probably is the general case with solar-type stars. This can be evidenced by normalizing Eq.\,(\ref{vorticity}) to dimensionless units. Measuring time in its viscous scale $R^2/\nu_{_{\rm T}}$ and multiplying Eq.\,(\ref{vorticity}) by this scale squared, gives the first and the second terms in the right-hand side of the normalised equation the coefficients of the Taylor (Ta) and Grashof (Gr) numbers
\begin{equation}
    {\rm Ta} = \frac{4\Omega^2 R^4}{\nu_{_{\rm T}}^2},\ \ \
    {\rm Gr} = \frac{gR^3}{\nu_{_{\rm T}}^2}\frac{\delta T}{T}
    \label{Ta_and_Gr}
\end{equation}
respectively, where $\delta T$ is the differential temperature. Direct numerical simulations \citep{Miesch_EA_06Diff_temperature} and mean-field models \citep{Kit_Ole_11DR} of the solar differential rotation give the value $\delta T/T \sim 10^{-5}$ for the normalised differential temperature varying moderately with depth. This leads to large characteristic values of ${\rm Gr} \sim {\rm Ta} \sim 10^7$ for the Sun. Each term in the left side of Eq.\,(\ref{vorticity}) scales to a much smaller value. There is no other way to satisfy this equation but the two terms on its right-hand side almost balance each other. This leads to the balance equation
\begin{equation}
    r\sin\theta\frac{\partial\Omega^2}{\partial z} -
    \frac{g}{r c_{\mathrm p}}\frac{\partial S}{\partial\theta} = 0 .
    \label{balance}
\end{equation}

Equation (\ref{vorticity}) shows that the meridional flow results from a slight deviation from the thermo-rotational balance of Eq.\,(\ref{balance}). The vorticity equation also informs on how the balance is maintained. Every term in the right-hand side of this equation alone can drive a meridional flow of order one kilometer per second \citep{Durney_96MF}. A considerable deviation from the balance would drive a fast meridional flow, which reacts back on the differential rotation and temperature to restore the balance. The meridional flow results from deviations from the thermo-rotational balance and also controls that the deviations are not large. This consideration shows that a reasonable model for the meridional flow alone is not possible. A realistic model has to solve consistently for the meridional flow, differential rotation and heat transport.

\begin{figure}
    \centering
    \includegraphics[width= 6 truecm]{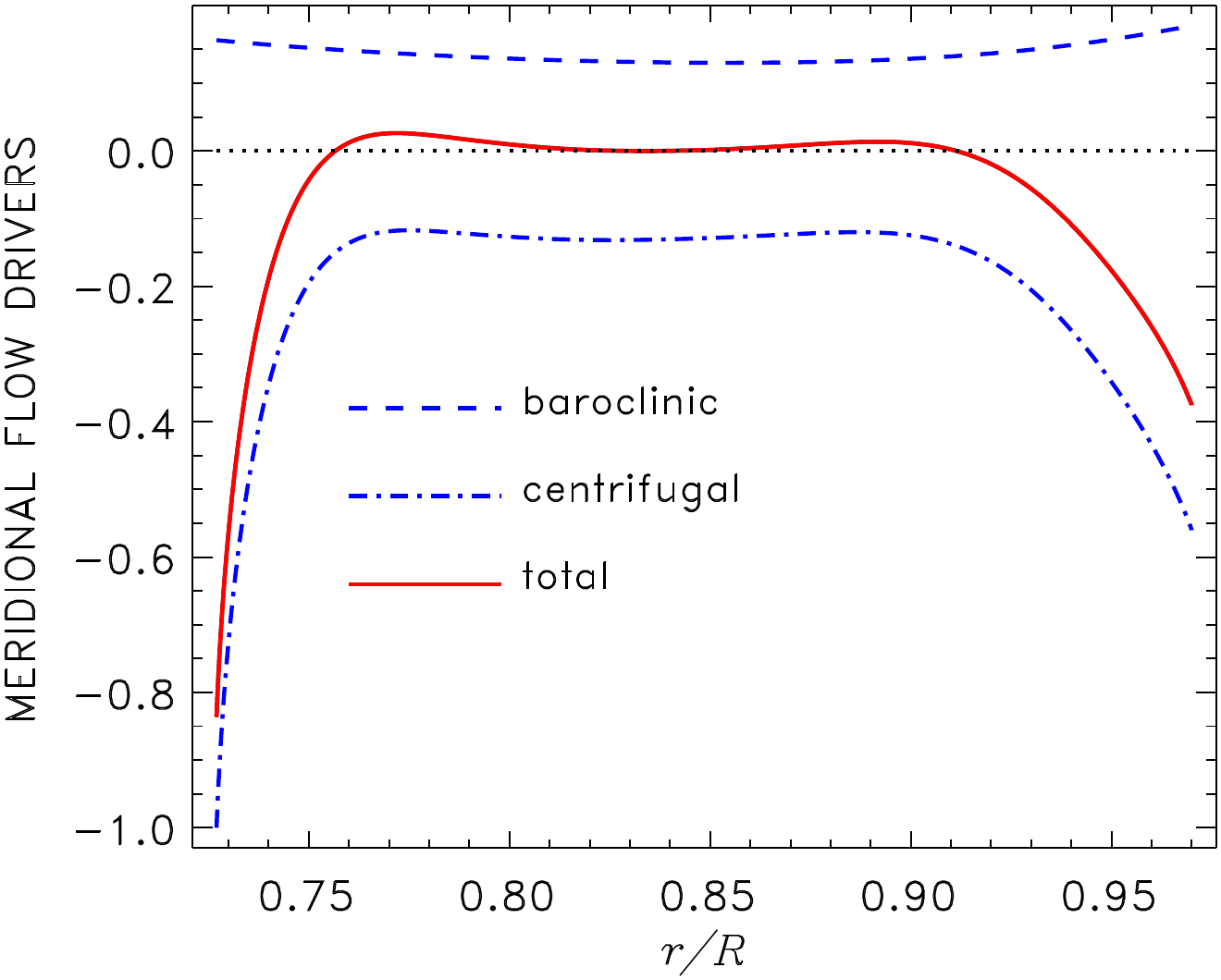}
    \caption{Depth profiles of the baroclinic and centrifugal driving
             terms of the meridional flow equation (\ref{vorticity}) for the 45$^\circ$ latitude computed with the mean-field model by \citet{Kit_Ole_11DR}. The driving terms are normalised to the maximum absolute value one and their sum is shown by the red line.}
    \label{drivers_1}
\end{figure}

Figure \ref{drivers_1} shows the depth profiles of the meridional flow drivers computed with a mean-field model. The sum of the baroclinic and centrifugal drivers is close to zero in the bulk of the convection zone, which is therefore close to the thermo-rotational balance of Eq.\,(\ref{balance}). The balance is however violated near the top and bottom boundaries. This is because of the stress-free boundary conditions employed in the model. This condition of zero surface density of external forces ensures that the meridional flow is controlled by \lq internal' processes inside the convection zone, not imposed externally. The stress-free condition together with zero radial velocity constitute a complete set of boundary conditions. The extra condition of the thermo-rotational balance cannot be satisfied near the boundary. Thin boundary layers form where the balance is violated. Deviation from the balance excites the meridional flow inside the boundary layers.

\begin{figure}
    \centering
    \includegraphics[width= 8 truecm]{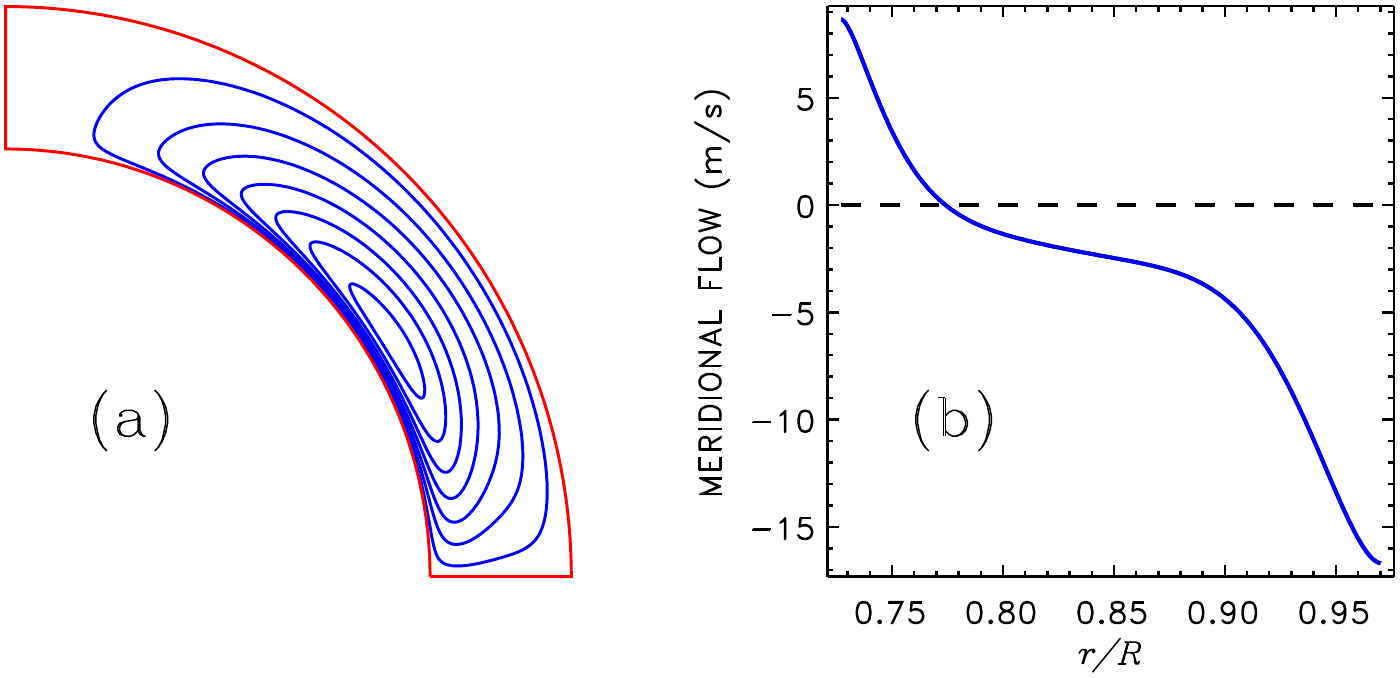}
    \caption{Meridional flow stream-lines (a) and depth profile
            of the meridional velocity for the 45$^\circ$ latitude (b)
            from the same mean-field model as Fig.\,\ref{drivers_1}.}
    \label{flow}
\end{figure}

The flow of Fig.\,\ref{flow} computed with the same mean-field model as Fig.\,\ref{drivers_1} attains its largest velocity on the boundaries and decreases inside the convection zone. The flow of this Figure is in at least qualitative agreement with the recent seismological detection    \citep{Rajaguru_Antia15,Gizon_EA_20Single_cell,Hanasoge_22LRSP}. An unsettled issue is how deep the equatorward meridional counter flow penetrates within the solar interior. Based on a kinematic dynamo modelling approach \cite{NC02} argued that a single cell flow penetrating below the convection zone in to the stable, overshoot layer is important for explaining the low-mid latitude appearance of sunspots. While subsequent arguments have been made both against and for such a possibility \citep{Gilman2004, Garaud2008}, we note that the most recent observations do not rule out a deep meridional counter flow \citep{Gizon_EA_20Single_cell}.

Since the convection cells become much smaller at the top of
the solar convection zone where the various scale heights are
much smaller compared to the interior, the nature of convection
clearly changes in a top layer and this complicates the issue
of a boundary layer there.  Observationally, helioseismic maps
of differential rotation show a near-surface shear layer at the
top of the solar convection zone. Recently \citet{Chou21a}
and \citet{Jha21}
have argued that this shear layer arises from the changed
nature of convection rather than from a violation of
Eq.\,(\ref{balance}).

The boundary layers in the solar model are not very thin (Fig.\,\ref{drivers_1}). Their thickness $D_{\rm E} \sim \sqrt{\nu_{_{\rm T}}/\Omega}$ decreases with rotation rate. For faster rotation, the meridional flow retreats to increasingly thin boundary layers and weakens inside the convection zone \citep{Kit_Ole_12DR}. Simultaneously, the differential rotation changes from the conical shape to cylinder-shaped pattern reflecting a faster increase of the Taylor number of Eq.\,(\ref{Ta_and_Gr}) with rotation rate compared to the Grashof number.
\subsection{Magnetic modifications}\label{sec3.2}
The meridional flow equation (Eq. (\ref{vorticity}) ) is modified with allowance for the large-scale axisymmetric magnetic field $\vec B$:
\begin{eqnarray}
    \frac{\partial\omega}{\partial t}
    &+& r\sin\theta\,{\vec\nabla}\cdot\left(\frac{{\vec V}\omega - {\vec V}_{\rm A}\omega_{\rm A}}{r\sin\theta}\right)
    + {\cal D}(\omega)
    \nonumber \\
    &=& r\sin\theta\frac{\partial(\Omega^2 - \Omega_{\rm A}^2)}{\partial z} -
    \frac{g}{r c_{\mathrm p}}\frac{\partial S}{\partial\theta}
    - \frac{g\rho}{2r\gamma P}\frac{\partial V_{\rm A}^2}{\partial\theta},
    \label{modified}
\end{eqnarray}
where meridional flow driving terms are again collected in the right-hand side of the equation. The magnetic terms in Eq.\,(\ref{modified}) are formulated in terms of the Alfven velocity ${\vec V}_{\rm A} = {\vec B}/\sqrt{\mu\rho}$ and the Alfven angular frequency $\Omega_{\rm A}$ for the toroidal field $B_\phi = \sqrt{\mu\rho}\,r\sin\theta\,\Omega_{\rm A}$; $\omega_{\rm A} = ({\vec\nabla}\times{\vec V}_{\rm A})_\phi$ is the magnetic vorticity and $\gamma = c_{\rm p}/c_{\rm v}$ is the adiabaticity index.

The first term on the right-hand side of Eq.\,(\ref{modified}) includes the non-conservative magnetic tension by the toroidal field. The minus sign in the contribution means that the tension force points towards the rotation axis, opposite to the centrifugal force. The last term on the right-hand side stands for the baroclinic driving by magnetic pressure. It was accounted for when deriving this term that the density varies much stronger in radius than in latitude and the convection zone stratification is close to adiabaticity.

It can be seen that magnetic contribution in the left-hand side of Eq.\,(\ref{modified}) includes the poloidal field only. This field is weak and this contribution is negligible for the sun. For stars with deep convection zones, the poloidal field can be strong \citep{Gregory_EA_12Strong_poloidal} and the magnetic advection of vorticity can be significant.

Assuming that the mean field in the deep convection zone of the sun is of order 1 Tesla, we could see that the magnetic terms in Eq.\,(\ref{modified}) are about two orders of magnitude smaller compared to the centrifugal driving. The magnetic terms are nevertheless large compared to each term on the left side of the equation. As in the hydrodynamical case, the meridional flow results from a disbalance of the driving terms in the right-hand side of Eq.\,(\ref{modified}) but the flow reacts back to ensure that the deviation from the balance remains small.

The magnetically modified thermo-rotational balance is global by nature. This in particular means that rotation law variation in torsional oscillations may not spatially coincide with the location of the magnetic fields producing the oscillations \citep{Pipin_Kosovichev_20TorsDyn}.

Magnetic field can also affect the meridional flow indirectly by modifying the differential temperature \citep{Spruit_03Barocline,Hanasoge_22LRSP} or differential rotation.

\section{Modelling the solar cycle: the paradigm shift
from the $\alpha \Omega$ dynamo to the flux transport dynamo}\label{sec4}

In the mean-field model, the magnetic field is assumed to be axisymmetric
and can be written as
\begin{equation}
\Bb = B_{\phi} (r, \theta, t) \, \ep + \nabla \times [A (r, \theta, t) \, \ep],
\label{mag}
\end{equation}
where $B_{\phi} (r, \theta) \, \ep$ is referred to as the toroidal field and
$\nabla \times [A (r, \theta) \, \ep] = \Bf_p$ gives the poloidal field.
The velocity field associated with large-scale flows can be written as
\begin{equation}
\Vb = \vf + r \sin \theta \, \, \Omega (r, \theta) \, {\bf e}_{\phi},
\label{vel}
\end{equation}
where $\Omega (r, \theta)$ is the angular velocity in the interior of the
Sun and $\vf$ is the meridional circulation having components
$V_r$ and $V_{\theta}$.  On substituting
Eq.\,(\ref{mag})and Eq.\,(\ref{vel}) into Eq.\,(\ref{induction})
with ${\vec{\cal E}}$ given by Eq.\,(\ref{EMF}), some
reasonable assumptions lead to the following coupled equations for the poloidal
and the toroidal fields
\begin{equation}
\frac{\pa A}{\pa t} + \frac{1}{s}(\vf.\nabla)(s A)
= \eta_T \left( \nabla^2 - \frac{1}{s^2} \right) A + \alpha B,
\label{eq:pol}
\end{equation}
\begin{equation}
\frac{\pa B}{\pa t}
+ \frac{1}{r} \left[ \frac{\pa}{\pa r}
(r V_r B) + \frac{\pa}{\pa \theta}(V_{\theta} B) \right]
= \eta_T \left( \nabla^2 - \frac{1}{s^2} \right) B
+ s(\Bf_p.\nabla)\Omega + \frac{1}{r}\frac{d\eta_T}{dr}
\frac{\partial}{\partial{r}}(r B),
\label{eq:tor}
\end{equation}
where $s = r \sin \theta$. Note that we are not including
the diamagnetic pumping term
in this discussion.

When the first efforts were made to construct mean field models of the solar
dynamo \citep{Parker55a, SKR66}, the existence of the meridional circulation was
not yet known. The early models which took $\vf = 0$ are now known as $\alpha \Omega$
dynamo models. In such models, the generation of the poloidal field involves
the $\alpha$-effect according to Eq.\,(\ref{eq:pol}) and the generation of the toroidal
field is due to differential rotation involving $\Omega$ according to Eq.\,(\ref{eq:tor}).
A remarkable result was that the $\alpha \Omega$
dynamo models could give periodic dynamo waves under certain
circumstances \citep{Parker55a, SK69}. This raised the possibility of explaining the
solar cycle with this model.  In order to model the butterfly diagram of sunspots,
we need to have the dynamo wave propagate in the equatorial direction.  The condition
for this was found to be
\begin{equation}
    \alpha \frac{\pa \Omega}{\pa r} < 0
    \label{PY}
\end{equation}
in the northern hemisphere of the Sun.  This is often referred to as the Parker--Yoshimura
sign rule \citep{Parker55a, Yoshimura_75}. In the 1970s when nothing was known about the
nature of the differential rotation underneath the solar surface, many models of the solar dynamo
were constructed by prescribing $\alpha$ and $\Omega$ in such a manner that the Eq.\,(\ref{PY})
was satisfied. Many of these models matched different aspects of the observational data of
solar cycles reasonably well and it seemed that the subject was progressing in the right
direction.

Several difficulties with the $\alpha \Omega$ dynamo models
started becoming apparent by the late 1980s.  Firstly,
as helioseismology started producing the first maps of
the angular velocity distribution inside the Sun, it was
found to be completely different from what was being
assumed in various $\alpha \Omega$ dynamo models. Secondly, it was established that the poloidal field of the Sun at
the surface propagates poleward with the progress of the
solar cycle, in contrast to the sunspots (forming from
the  toroidal field) which appear closer to the
equator as the cycle progresses
\citep{Wang89}.  In the simplest kinds of $\alpha \Omega$
dynamo models without meridional circulation, the poloidal
and toroidal fields remain coupled to each other, and it
is not possible to make them move in opposite directions.
Thirdly and lastly, simulations of sunspot formation
indicated that the toroidal field must be  much stronger
than what used to be assumed.  Bipolar sunspots form
when parts of the toroidal field rise through the convection
zone due to magnetic buoyancy \citep{Parker55b}. Detailed
simulations of this process based on the thin flux tube
equation \citep{Spruit81,Chou90} showed that the Coriolis force
due to the solar rotation tries to divert the rising
flux tubes towards high latitudes
\citep{CG87, Chou89}. Only if the magnetic
field inside the flux tubes is sufficiently strong, it is
able to counter the Coriolis force in such a manner
that there is a match with the observational data
\citep{DSilva93, Fan93, Cali95}.

The flux transport dynamo model arose in response
to these difficulties with the $\alpha \Omega$
dynamo models. Bipolar sunspot pairs on the solar
surface appear with a tilt \citep{Hale19}---due to the action
of the Coriolis force \citep{DSilva93}. \citet{Bab61}
and \citet{Lei69} realized that the decay of such
tilted bipolar sunspot pairs would give rise to a
poloidal field.  The flux transport dynamo model
invokes this Babcock--Leighton mechanism for the
generation of the poloidal field. Unlike the canonical $\alpha$-effect,
this mechanism does not require the toroidal field to be
sufficiently weak. However, when the
Babcock--Leighton mechanism is combined with the
differential rotation given
by helioseismology in a minimalistic dynamo model,
the Parker--Yoshimura condition given by Eq.\,(\ref{PY})
is not satisfied at the low latitudes and dynamo waves
are found to propagate in the poleward direction
implying that sunspots would appear at higher latitudes
with the progress of the solar cycle \citep{CSD95}, in
contradiction with observations.  We certainly need
something else to turn things around.  The meridional
circulation was the first proposition which provides a way out of this conundrum.

\begin{figure}
    \centering
   \includegraphics[width= 0.8\textwidth]{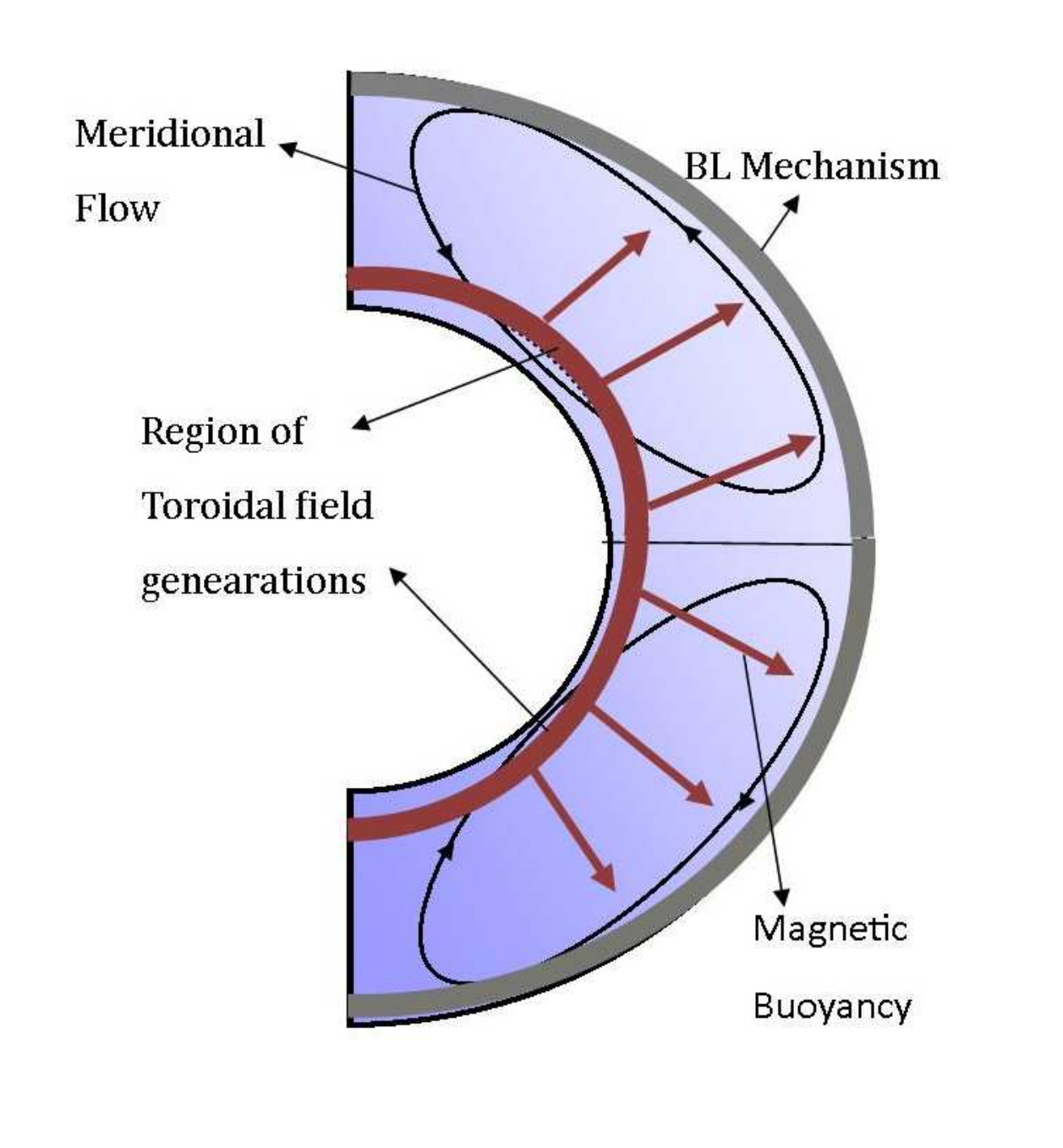}
    \caption{A cartoon indicating the essential ingredients of
    the flux transport dynamo model. Taken from the PhD thesis \citet{Hazra_thesis_2018}.}
    \label{FTD}
\end{figure}

Fig.\,\ref{FTD} is a cartoon summarizing how the flux transport dynamo
model works.  The dark red region at the bottom of the convection
zone is where helioseismology has discovered a strong layer of
differential rotation which overlaps with the stable overshoot layer beneath the convection zone. Dynamo models which incorporate direct helioseismic observations indicate that the toroidal field begins to be inducted in the convection zone \citep{munoz09} and is subsequently amplified, stored and transported equatorward in the tachocline region \citep{NC02}. Toroidal field that escapes out of this layer in to convection zone rises to form sunspots due to magnetic
buoyancy indicated by the dark red arrows. The decay of sunspots near the
surface indicated by the greyish color gives rise to the poloidal
field by the Babcock--Leighton mechanism.  The meridional circulation
is shown by the black contours.  It is equatorward at the bottom of
the convection zone so that the toroidal field generated there is
advected equatorward, producing sunspots closer to the equator with
the progress of the solar cycle.  On the other hand, the \MC\ is
poleward near the surface so that the poloidal field generated there
is advected poleward.

The first 2D axisymmetric models of the flux transport dynamo were
constructed in the mid-1990s \citep{CSD95, Durney_95}, although some of the basic ideas were put forth on the basis of a 1D model in an
earlier paper by \citet{Wang91}. That the meridional circulation
can reverse the direction of the dynamo wave was demonstrated
convincingly by \citet{CSD95} and paved the way for the formulation
of the flux transport dynamo model.  Within the next few years,
different groups studied different aspects of the model
\citep{Dur97,DC99, NC01,  KRS01, NC02,Bon02,Gue04, CNC04, Chou04}.
This model could explain various aspects of observational data
pertaining to the solar cycle, especially the butterfly diagram
of sunspots along with the time-latitude distribution of the
poloidal field at the surface \citep{CNC04}.

A majority of the flux transport dynamo calculations
assumed such a single-cell meridional flow as shown in Fig.\,\ref{FTD}. The nature of the \MC\ deeper down in the convection zone remained uncertain till fairly recently and some groups claimed a more complicated, multi-cellular profile \citep{Zhao13}. Subsequent research has shown that under certain circumstances the flux transport paradigm can work even in the presence of complex, multi-cellular meridional flow \citep{HKC14,hazra2016}. However, helioseismology results from different groups are now converging on a single-cell flow pattern \citep{Rajaguru_Antia15, Gizon_EA_20Single_cell}
in agreement with what had been assumed in the majority of flux transport dynamo calculations, triumphantly validating the flux transport dynamo model.

The period of the flux transport dynamo is essentially set
by the time scale of the \MC. When other parameters are held fixed,
the period $T$ and the amplitude $v_0$ of the \MC\ are found
to obey the approximate relation
\begin{equation}\label{eq:mc-period}
    T \propto v_0^{-\gamma}.
\end{equation}
The index $\gamma$ is found to have a value close to 1 in
different models of the flux transport dynamo \citep{DC99, yeates2008}.

We note as a caveat that flux transport dynamo models incorporating both radial and latitudinal turbulent pumping as gleaned from magnetoconvection simulations can explain many of the observed features of the solar cycle, even in the absence of meridional circulation -- circumventing the constraint of the Parker-Yoshimura sign rule \citep{Hazra2019}. However, unlike meridional circulation, the turbulent pumping profile in the solar convection zone remains completely unconstrained by observations.

One limitation of the 2D axisymmetric models of the flux transport dynamo
is that the Babcock--Leighton mechanism is intrinsically a 3D mechanism
and can be treated in 2D models only by making drastically simplifying
assumptions.  In fact, there has been a debate about the best way of
treating this mechanism in 2D models \citep{Dur97, NC01, Muniz10}. One possible approach of handling this mechanism more realistically is to develop 3D
kinematic models in which the magnetic field is treated in 3D so that the dynamics of tilted bipolar sunspots can be computed explicitly \citep{Yeates13,Miesch14, HCM17, HM18}. An important recent development in this context is a 3D kinematic Babcock--Leighton flux transport dynamo where the buoyant emergence of flux tubes is treated as a dynamic, magnetic field dependent process in a self-consistent manner \citep{kumar2019}. All the recent developments in the 3D kinematic dynamo models are reviewed by \citet{Hazra2021_review}.

\section{Extrapolation to stellar dynamos}\label{sec5}

Magnetic field and sun-like magnetic cycle have been observed in many solar-type stars with outer convection zone \citep[e.g.,][]{Wilson78, Noyes84a, Baliunas95, Donati97}. Unlike the Sun, for which we have a lot of detailed observational data available, observation of surface magnetic field for other stars is quite limited. The observational estimate of magnetic activity for other stars mostly comes from indirect proxies of the magnetic field such as measurements of chromospheric Ca II H \& K lines \citep{Wilson78, Noyes84a, Baliunas95} and Coronal X-ray emission \citep{wright11, WD16}. Also, one of the major difficulties in measuring stellar activity is that we need a long-term programme for monitoring stars as their cycle period will likely to be commensurate with the 11-year solar cycle period. Thanks to Mount Wilson observatory monitoring program \cite{Wilson78}, we have long-term data of Ca II H \& K flux for 111 stars from spectral type F2-M2 on or near main sequence. Using this data, \citet{Noyes84a} found that the magnetic activity of stars increases with the rotation rate. Actually, the magnetic activity  better correlates with
Rossby number, which is a ratio of the rotation period to the convective turnover time. In Figure~8 of \citet{Noyes84a}, it is shown how the magnetic activity varies with the Rossby number. The magnetic activity first increases rapidly
with increasing rotation rate (or decreasing Rossby number), and then it increases very slowly or even seems to be independent of Rossby number for rapidly rotating stars. This result was corroborated by other independent studies from coronal X-ray emission \citep[e.g.,][]{Hempelmann95, wright11}. Recently Zeeman Doppler Imaging (ZDI) technique \citep{Donati97} emerges as a promising way of reconstructing surface magnetic field from other stars. Using this method, \citet{Vidotto14} analysed 73 late-F, G, K and M dwarf stars and reported a similar rotation-activity relation.

\begin{figure*}
    \centering
    \includegraphics[width=0.465\textwidth]{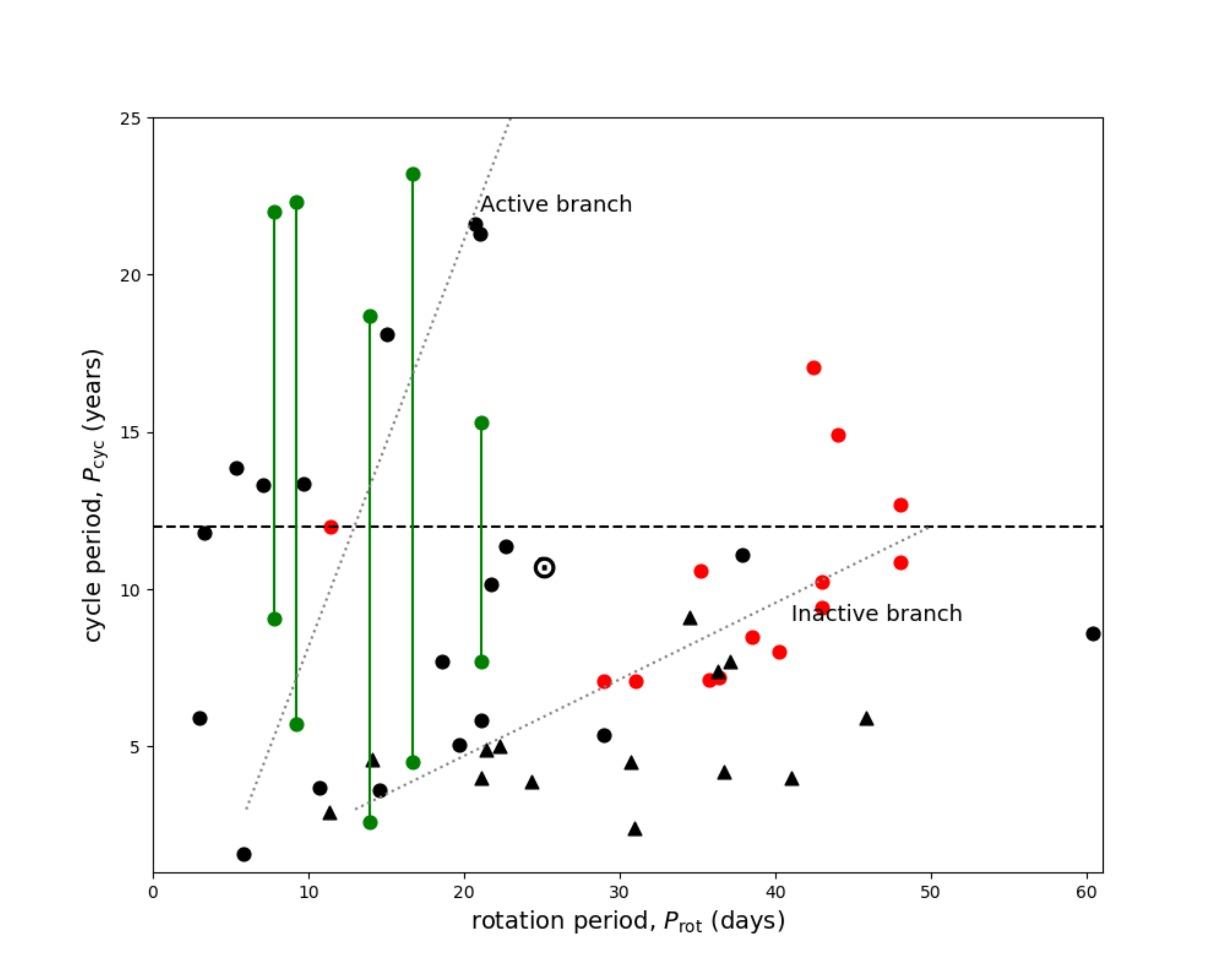}
    \includegraphics[width=0.49\textwidth]{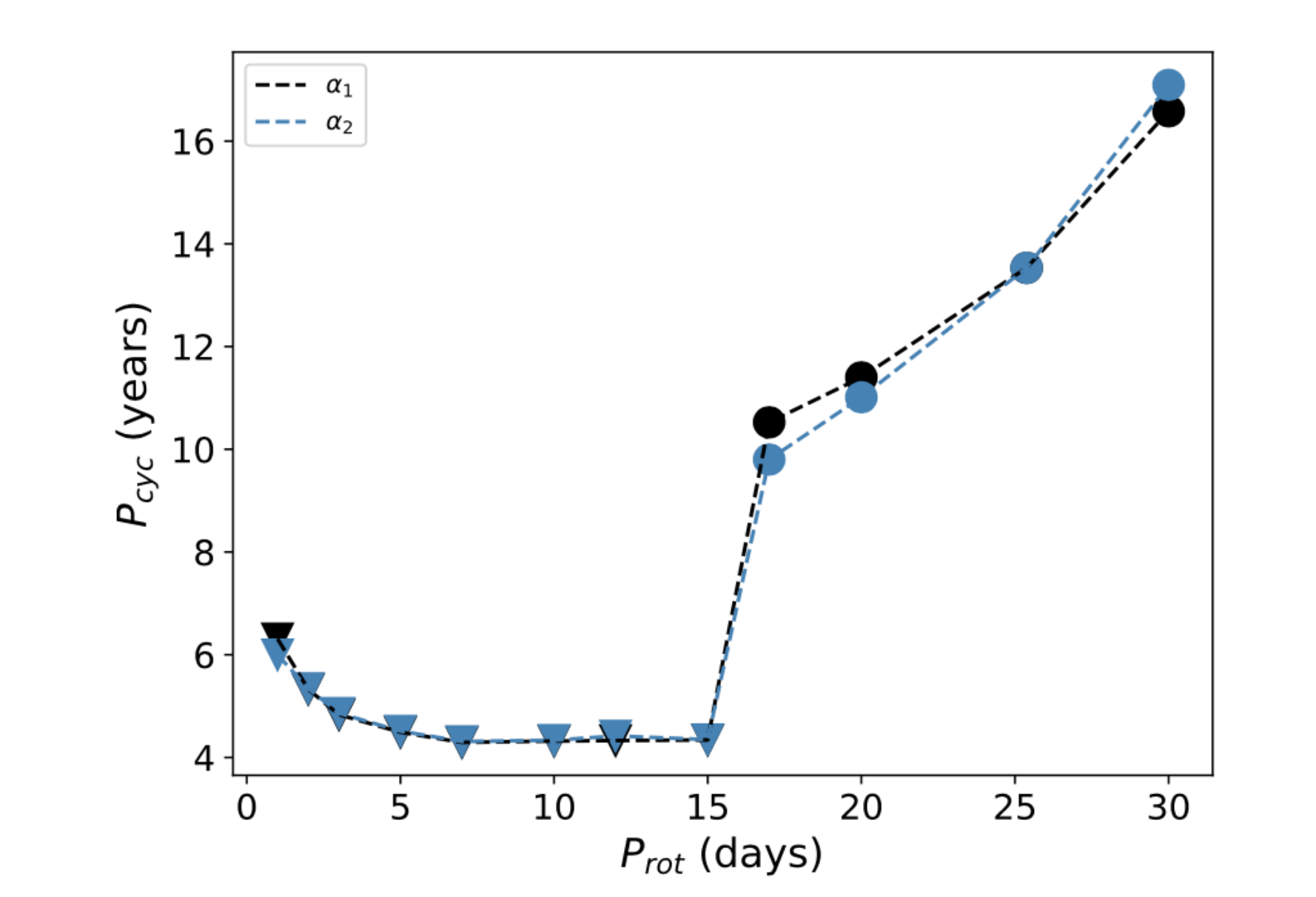}
    \caption{Left: Rotational dependency of cycle period from observations - P$_{\rm cyc}$ vs P$_{\rm rot}$ plot (Taken from \citet{BoroSaikia18}). The red symbols are stars with well defined activity cycles, green symbols are stars with multiple activity cycles, and black symbols are for stars with unconfirmed activity cycles. Mount Wilson stars are denoted as filled circles and triangles represent HARPS stars. The active and inactive branches from \citet{Bohm07} are shown in dashed black lines. The Sun is shown as $\odot$. Right: P$_{\rm cyc}$ vs P$_{\rm rot}$ plot from theoretical model of \citet{Hazra2019}. The black and blue colors represent two types of treatment in their Babcock-Leighton $\alpha$ effect. Stars with quadrupolar and dipolar parities  are shown in triangular and circular symbols respectively.}
    \label{fig:pcyc-prot}
\end{figure*}

While stellar activity follows a clear dependency on the rotation rate of the stars, the dependence of the stellar cycle period on rotation is somewhat complicated \citep{VP80, Noyes84b}. Mount Wilson sample of Ca II H \& K shows two distinct branches the active, young one and old slowly rotating one with a gap between them known as Vaughan \& Preston gap \citep{VP80}. It has been found that the cycle period decreases with decreasing rotation period of the stars in both of the branches. This data was further analysed carefully by many others \citep{SB99, Saar2002, Bohm07} reporting similar trends. Figure~1 from \citet{Bohm07} shows the clear trend of decreasing cycle period with increasing rotation rate along the inactive and active branches. A recent analysis of a larger sample (4454 Cool stars) shows that the Vaughan and Preston gap might be a result of a lack of data in the Wilson sample \citep{BoroSaikia18}. The left panel of Fig.~\ref{fig:pcyc-prot} (taken from \citet{BoroSaikia18}) shows the P$_{\rm cyc}$-P$_{\rm rot}$ diagram for the stars with observed cycle period. The stars with well-defined cycles in their sample show an increasing trend of cycle period with rotation period and there is no clear gap between inactive and active branches of stars. However, as it is clear from the figure, the uncertainty lies in the fast-rotating active branch. Similar results are also reported by \citet{Olspert2018} by an individual probabilistic analysis of the Ca II H \& K data.

Many theoretical efforts have been made to understand the relation
of magnetic activity and cycle period with rotation period of the stars \citep[e.g.,][]{DR1982, Robinson82, Brandenburg94, Nandy04, KO15, Jouve10, KKC14, Strugarek2017, Warnecke2018, Hazra2019}. There were some early efforts from traditional $\alpha\Omega$ mean-field dynamo \citep{DR1982, Robinson82, Brandenburg94}, before the importance of meridional circulation was properly recognized in dynamo theory (See section~4 for details), to understand the observational behavior of stars. The observed dependence of magnetic activity on the rotation rate of the star is naturally explained from the mean-field $\alpha\Omega$ dynamo theory. The $\Omega$-effect directly depends on the differential rotation which is directly connected to the rotation rate of the star. Also, the $\alpha$-effect which is a measure of helical turbulence naturally relates to the rotation rate of the star. For kinematic dynamo, in the linear regime, the dynamo can sustain if the dynamo number D = $\frac{\alpha R^3}{\eta} \frac{1}{r}\frac{\partial \Omega}{\partial r}$ ($\eta$ is the co-efficient of turbulent diffusivity and R is the outer stellar radius) exceeds a critical value $D_c$. In that case, the period of the dynamo cycle $P_{\rm cyc}$ $\propto$ D$^{-1/2}$ \citep{Noyes84b}. Hence $P_{\rm cyc} \propto \Omega^{-1}$, which is in agreement with the stellar observation. However, in the non-linear regime, where the magnetic field grows until the Lorentz force alters the velocity field to permit some equilibrium, the $\alpha$-effect or velocity shear is reduced as the field strength increases. As a result, the dynamo number gets reduced until a steady state is achieved and the cycle period has the approximately same value as it had for $D= D_c$. The quenching of dynamo action gives a cycle period almost independent of rotation $\Omega$.

Meanwhile, the importance of meridional circulation in the solar dynamo theory, hence the Flux Transport Dynamo (FTD) theory (see Section-4 for details) was established to explain many properties of the solar magnetic field \citep{CSD95, CNC04}. The first comprehensive model of FTD for solar-like stars was carried out by \citet{Jouve10}. Two main ingredients of the FTD model differential rotation and meridional circulation were obtained from 3D hydrodynamic simulations as the observational data for them is not available for other stars. The 3D hydrodynamic simulations result a slower meridional circulation with an increasing rotation rate. In FTD models, as the cycle period is inversely proportional to the speed of meridional circulation \citep{DC99, CNC04}, the computed cycle periods with different rotation rates from these models are not compatible with observations. Similar results were reported from the scaling relation of stellar dynamo \citep{Nandy04}. \citet{KKC14} also constructed a theoretical model for stellar dynamo based on FTD model. They used the differential rotation and meridional circulation for stars with rotation periods of 1 day to 30 days from mean-field hydrodynamic models as presented in Section 2. They also reported an increase in the cycle period with increasing rotation rate, as the amplitude of meridional circulation from the mean-field hydrodynamic model decreases with the increasing rotation rate of stars.

Recently, \citet{Hazra2019} extended the study of \citet{KKC14} by incorporating radial turbulent pumping. Turbulent pumping was found to be unavoidable in a stratified stellar convection zone due to the topological asymmetric convective flows \citep{Tobias_EA_98Dia_by_DNS, Kapyla06, MH11}. A few previous studies in a solar context already showed that pumping is important in transporting poloidal field from the surface to the deeper convection zone and to match the results of FTD models with observed surface magnetic field \citep{Guerrero08, Cameron12, KN12, KC16, hazra2016}. The inclusion of turbulent pumping suppresses the diffusion of the horizontal field and makes the behavior dynamo different than the traditional flux transport dynamo model. In addition to explaining the increasing magnetic activity with the rotation, the model of \citet{Hazra2019} can explain the decreasing trend of the cycle period with the increasing rotation rate of stars for the inactive branch of slowly rotating stars. In the right panel of Fig.~\ref{fig:pcyc-prot}, the dependence of cycle period with the rotation period of the stars is shown for two types of treatment of Babcock-Leighton $\alpha$ effect with rotation (see section 2.3 in \citet{Hazra2019} for details). A direct comparison of their result (right panel of Fig.~\ref{fig:pcyc-prot} ) with the observed trend of cycle period dependency on rotation (left panel of Fig.~\ref{fig:pcyc-prot}) makes it clear that the observed trend is reproduced qualitatively well but the active branch is still needed careful further study. They also reported that the global magnetic field changes from dipolar to quadrupolar parity in rapidly rotating stars for a rotation period of less than 17 days. The global magnetic field distribution in the Sun and stars, its parity, and the structure of coronal magnetic fields are governed by the dynamo mechanism \citep{dash2023} and surface emergence and evolution of magnetic flux \citep{nandy2018, Kavanagh2021}. The magnetic field topology in turn determines the global stellar magnetosphere and magnetized stellar wind \citep{Reville2015,Vidotto2014a} that play critical roles in star-planet interactions \citep{das2019,basak2021,Carolan_Hazra_2021} and the forcing of (exo)planetary space environments \citep{nandy2021b, Hazra2022}. Also, the stellar magnetic cycle alters the total X-ray and EUV (XUV) radiation from host stars affecting exoplanetary atmospheres \citep{Hazra2020_exo}.

There is another idea that the observed dependency of stellar activity cycles on rotation rates might be a manifestation of the dependence on the effective temperature of stars \citep{Kit2023}. By combining models of differential rotation and dynamo together for stars with different masses, \citet{Kit2023} found shorter cycles for hotter stars. Also, note that the hotter stars rotate faster on average. Hence computed shorter cycles for hotter stars are basically for fast rotators.

The flux transport dynamo paradigm can in fact be elegantly captured via a mathematical formulation based on time-delay differential equations \citep{wilmotsmith2006}. \cite{tripathi2021} show that such a truncated Babcock--Leighton model imbibing the effects of fluctuations and noise can simultaneously explain the observed bimodal distribution of long-term sunspot time series, the breakdown of gyrochronology relations in middle aged solar-type stars and the relative low activity of the Sun compared to other Sun-like stars. This lends further credence to the philosophy that the basic ideas of flux transport dynamo theory gleaned in the context of the Sun may apply to other solar-like stars and across a substantial phase of solar evolution.

\section{Temporal variations of the meridional circulation and solar cycle fluctuations}\label{sec6}

Since the period of the flux transport dynamo depends on the strength of the \MC, as indicated in Eq. (\ref{eq:mc-period}), it is obvious that fluctuations in the \MC\ would have an effect on the dynamo.  We now discuss what we know about the temporal variations of the \MC\ and how they may affect the dynamo.

\subsection{Evidence for \MC\ variations}
\begin{figure*}
    \centering
    \includegraphics[width=0.9\textwidth]{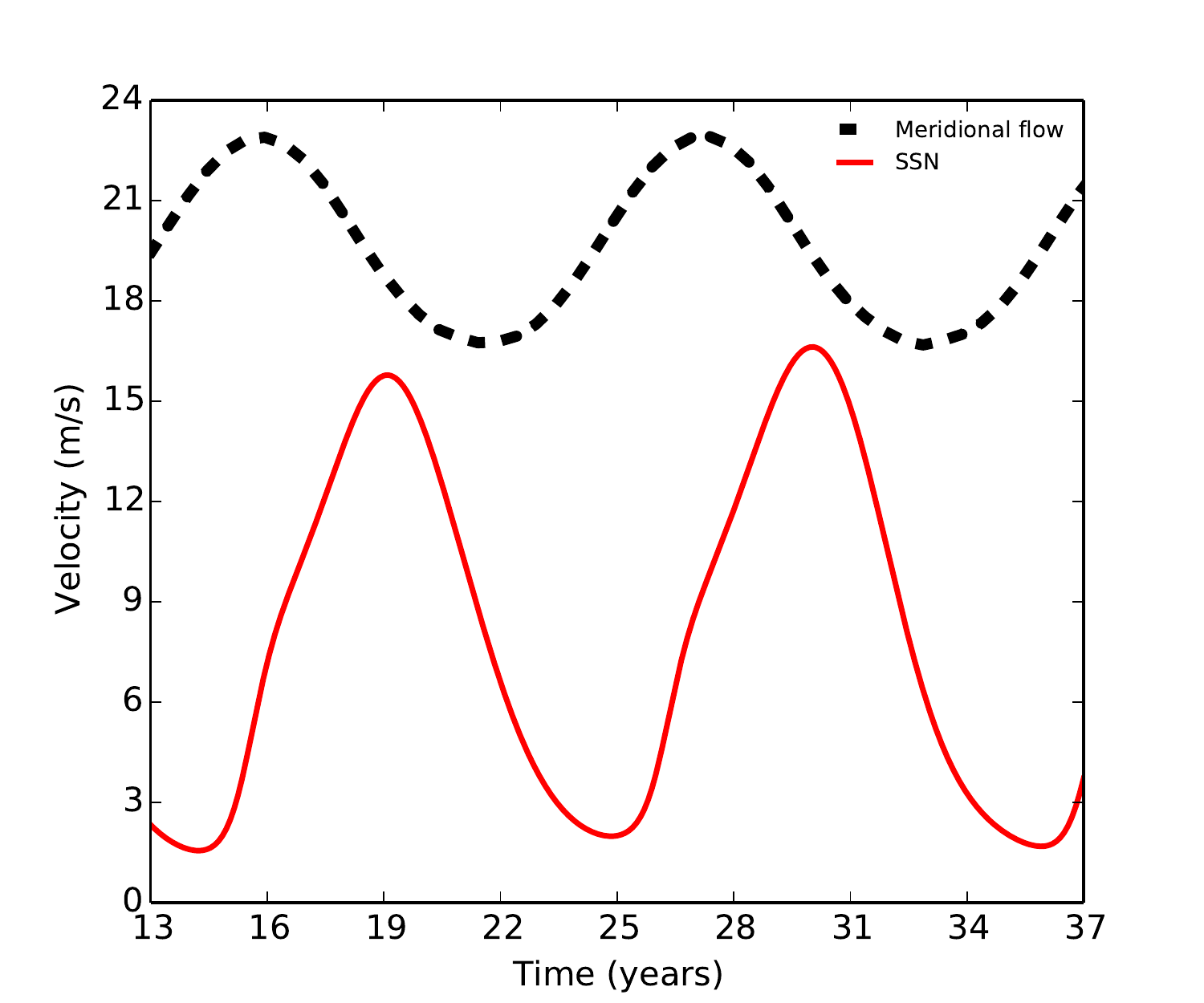}
    \caption{Theoretical estimate of variation of meridional circulation with the solar cycle. The black dashed line shows the amplitude of meridional circulation and the red solid line shows two synthetic solar cycles.
    Adapted from \citet{HC17}.}
    \label{fig:mc_periodic}
\end{figure*}

A variation of the \MC\ with the solar cycle has been inferred both from helioseismology \citep[e.g.,][]{ChouDai2001, Beck2002, BA2010, Komm2015} and from the tracking of surface markers \citep{Hathaway2010, Mahajan2021}. It has been found that the \MC\ becomes weaker at the time of sunspot maximum. From GONG full-disk Dopplergrams and HMI instrument on SDO, \citet{Komm2015} computed the temporal variation of the amplitude of meridional circulation near the surface at three depths of 2.0 Mm, 7.1 Mm and 11.6 Mm over latitudes as shown in their figure~9. It is clear from the figure that the amplitude of the meridional flow became weaker near solar maxima around the years 2002 and 2014.
This is presumably caused by the back-reaction of the dynamo-generated magnetic field on the large-scale flows. Some effects of the cyclic variation of the \MC\ can be studied by introducing a simple quenching by the magnetic field \citep{Karak12}. However, a proper theoretical understanding
requires the solving of Eq.~(\ref{vorticity}) simultaneously with the dynamo equations (Eq.~(\ref{eq:pol}) and Eq.~(\ref{eq:tor})). \citet{HC17} developed a perturbation approach to study this problem.  Their result is shown in Fig.~\ref{fig:mc_periodic} compares favorably with the observational data. In this context, an independent study by \citet{nandy2011} claims that a relatively faster meridional flow in the rising phase of the cycle followed by a slower flow in the declining phase (on average throughout the meridional flow loop in the convection zone) can explain the occurrence of unusually deep minima between solar activity cycles; the results of \cite{HC17} does not appear to be inconsistent with this finding. Another numerical study by \cite{saha2022} - based on a flux transport dynamo model - indicates that meridional circulation can reproduce the
(observed) cyclic modulation of weakened activity during grand minima phases such as the Maunder minimum. However, very long-term observations of surface flow variations do not exist and current capabilities do not allow setting strong constraints on deep flow variations.

It may be noted that the back-reaction of the magnetic field also causes periodic variations in the differential rotation, the so-called torsional oscillations.  There have been efforts to model this also within the framework of the flux transport dynamo \citep[e.g.,][]{CCC09}.

\begin{figure*}
    \centering
    \includegraphics[width=1.0\textwidth,angle=0,origin=c]{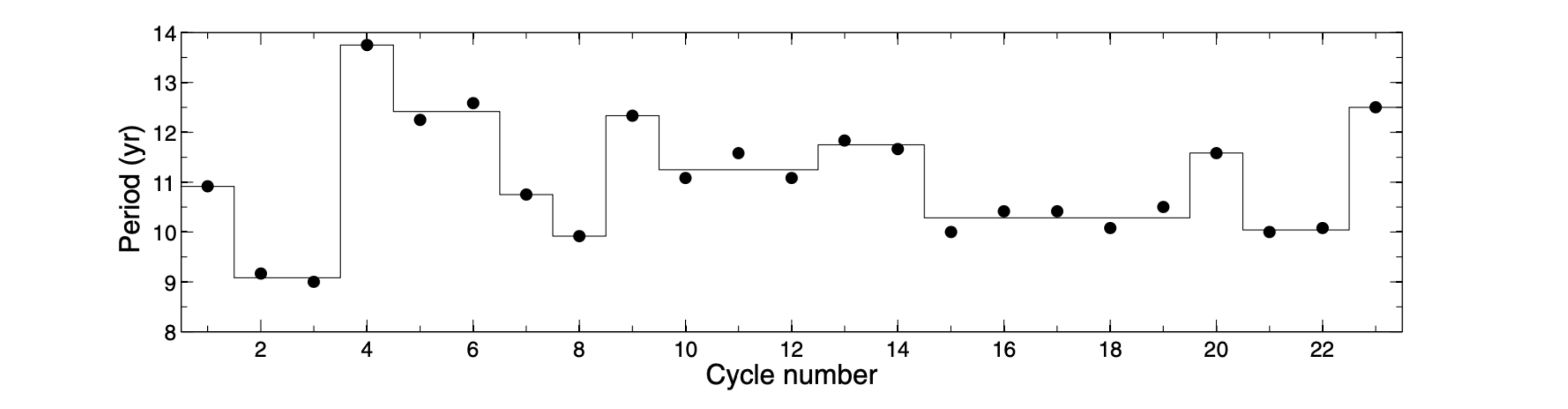}
    \caption{Durations of various solar cycles beginning with the solar cycle 1. The solid filled circles show the observed period of  the last 23 cycles. The solid line is for guiding the eye to discern the patterns in the variations
    of the solar cycle durations. Adapted from \citet{Karak2011}.}
    \label{fig:mc_random}
\end{figure*}

We are interested here in the question of whether there are more random, non-periodic variations in the \MC.  Since we have reliable observational data about the \MC\ only for about a quarter century, this question cannot be answered on the basis of direct observations.  However, as the periods of the cycles depend
on the strength of the \MC, we may use the data about the durations of past cycles to draw inferences about the variations of the \MC\ \citep{Karak2011}. In Fig.~\ref{fig:mc_random} by plotting the durations of various past cycles, we see that cycles 10 to 14 had an almost constant period somewhat longer than 11 yr, suggesting that the meridional circulation was probably weaker in that era. Then cycles 15 to 19 had an almost constant period somewhat shorter than 11 yr, suggesting a stronger meridional circulation at that time. Based on such considerations, \citet{Karak2011} concluded that the meridional circulation had some random fluctuations with a coherence time of a few decades---perhaps in the range between 20 and 50 yr. Such fluctuations are expected to be a major cause behind the irregularities of the solar cycle.

\subsection{Possible causes behind the irregularities of the solar cycle}
The earliest idea for explaining solar cycle irregularities was that this is a manifestation of nonlinear
chaos \citep{Weiss1984}. Although the dynamo process certainly involves various kinds of nonlinearities, the most obvious nonlinearities are found not to produce any sustained chaotic behaviour and the various random fluctuations associated with the dynamo may be the more likely candidates for producing
the cycle irregularities \citep{Chou92}. However, there is one kind of observation that is presumably a signature of chaos: the Gnevyshev-Ohl effect obeyed over many cycles that the even cycle was stronger than the previous odd cycle. This is presumably due to period doubling just beyond bifurcation \citep{Charbonneau05}.

We now try to identify the possible sources of fluctuations in the flux transport dynamo model. The Babcock-Leighton process depends on the tilts of active regions. We see a scatter in the tilt angles \citep{SK12}, presumably caused by the turbulent buffeting of flux tubes rising through the convection zone \citep{Longcope02}. \citet{CCJ07} proposed that the scatter in tilts gives rise to random fluctuations in the Babcock-Leighton process. This idea enabled them to make the first successful dynamo-based prediction of a solar cycle \citep{CCJ07, Jiang07}. More support for this idea has come from observational data \citep{DasiEspuig10} and simulations \citep{KM17}.
Fluctuations in the Babcock--Leighton process have also been invoked
to model the hemispheric asymmetry of sunspot cycles \citep{Goel09} and
the Maunder minimum \citep{Karak09}.
Stochastic fluctuations in the source terms for the poloidal field, both in the context of the mean-field $\alpha$-effect and the Babcock--Leighton mechanism have been utilized within the flux transport dynamo paradigm to demonstrate the importance of these fluctuations in the occurrence and recovery from grand minima episodes \citep{hazra2014,passos2014} and in the genesis of hemispheric decoupling and parity modulation in the sunspot cycle \citep{hazranandy2019}.

One major limitation of using fluctuations in the Babcock-Leighton process alone for explaining the irregularities in the solar cycle is that these fluctuations cannot produce much variations in cycle durations \citep[see however][]{Kit2018}. To explain the observed variations in the cycle periods, we need something else like the fluctuations in the meridional circulation or other transport coefficients such as turbulent pumping. We now turn to a discussion of the effects that such fluctuations would produce on the dynamo.

\subsection{The effects of random fluctuations in the \MC}
\citet{Karak10} varied the \MC\ to match the periods of various solar cycles in the twentieth century and found that
even the amplitudes of the cycle got matched to a certain extent.   This was a clear indication that one of the causes
behind the irregularities of the solar cycle was the fluctuations in the \MC.

Suppose the meridional circulation has slowed down due to fluctuations, which will make the cycles longer.
Diffusion will have more time to act and will try to make the cycles weaker. This would cause an anti-correlation
between the strength of the cycle and its duration.
A consequence of stronger cycles having shorter duration is that they should rise faster. The anti-correlation between
the rise time and the cycle strength has been known for a long time and is called the Waldmeier effect. \citet{Karak2011} succeeded in explaining the Waldmeier effect by incorporating fluctuations in the \MC\ in their dynamo model.

\citet{CK12} developed a comprehensive model of grand minima by including fluctuations in both the Babcock-Leighton process and in the meridional circulation in their dynamo simulations. By analyzing polar ice cores ($^{14}$C data), \citet{Usoskin2007} arrived at the result that there were about 27 grand minima in the last 11,000 yr. The results of \citet{CK12} are in broad agreement with this.

\subsection{Solar cycle fluctuations and cycle forecasts}

Can we utilize our understanding of solar cycle fluctuations to predict future sunspot cycles. Observations indicate that the polar field is a good precursor of the following sunspot cycle and a dynamo basis for this was already alluded to early on \citep{schatten1978}. The first suggestion of using dynamo models for forecasting future solar cycle amplitudes -- using poloidal field as inputs --- was made by \cite{nandy2002}. Subsequently detailed models based on the flux transport paradigm were worked out. It is noteworthy that while a variety of prediction techniques exist in the literature \citep{petrovay2020}, predictions based on the Babcock--Leighton paradigm and data driven flux transport dynamo models appear to now provide consistent results \citep{Nandy2021}.

The theoretical explanation of why the polar field is such a good precursor was provided by \citet{Jiang07}, who pointed out that the polar field captures the essential outcome of the fluctuations in the Babcock-Leighton process. A series of papers utilizing the flux transport paradigm and stochastic fluctuations in the Babcock--Leighton source term
 -- established the importance of cycle memory, i.e., the propagation of information of past polar fields to future sunspot cycles \citep{yeates2008,KN12}. However, these studies did not look at aspects related to meridional flow variations.

If fluctuations in the meridional circulation are also important, can we find a precursor to capture the effects of that? It turns out
that there is a time lag between the meridional circulation and its effect on the dynamo.  The strength of a cycle
does not depend on the value of the meridional circulation at the cycle maximum, but on its value a few years
earlier---when the previous cycle was decaying.  As a result, the decay rate of the previous cycle has a correlation
with the strength of the next cycle and provides the appropriate precursor which encapsulates the effect of fluctuations
in the \MC\ \citep{HKBC15}.

\begin{figure*}[t]
    \centering
    \includegraphics[width=0.99\textwidth]{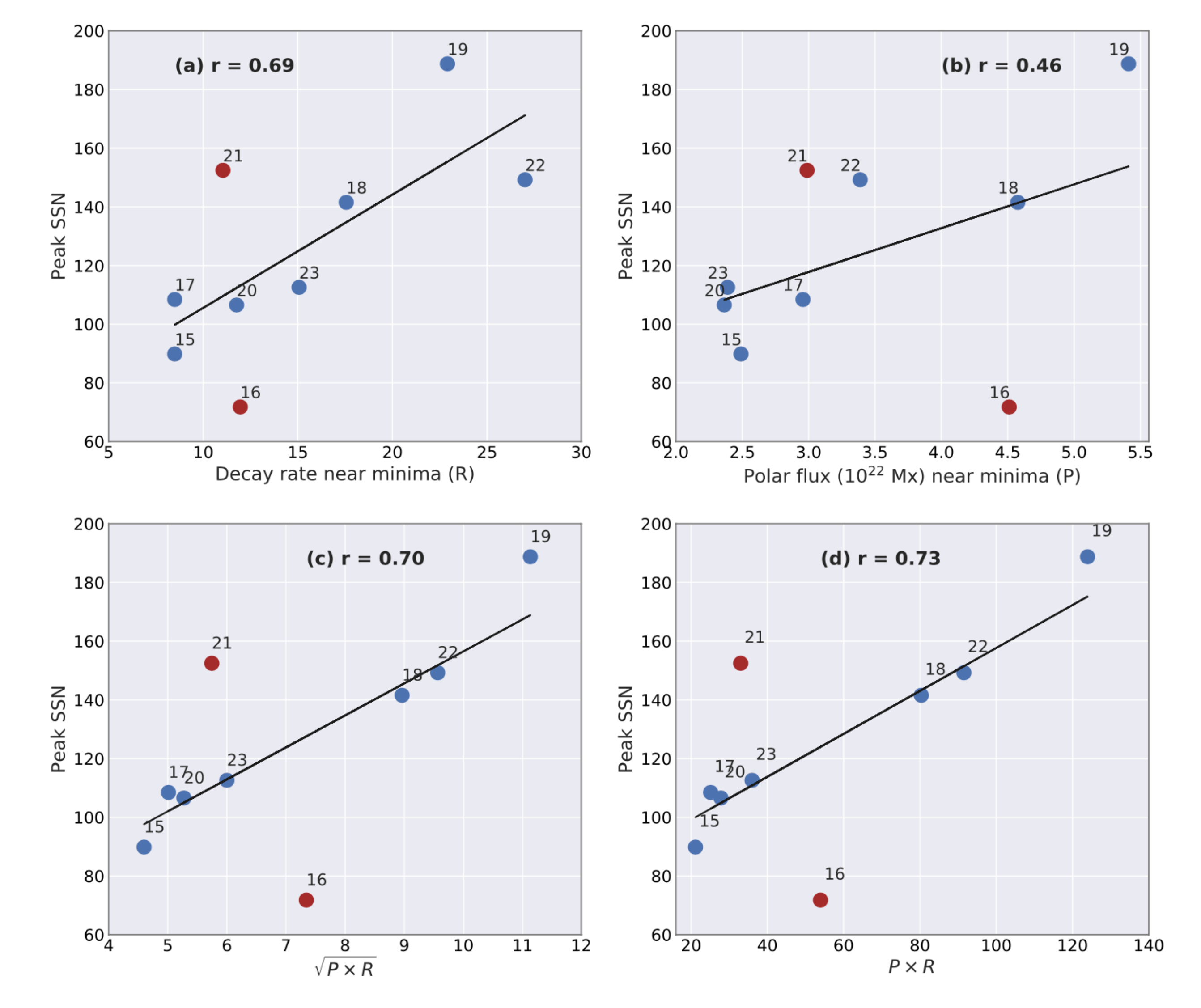}
    \caption{The correlation plots of various precursors with the amplitude of the next cycle. Correlations of next cycle amplitude with (a) the decay rate at the late phase of the cycle $R$, (b) the polar field near minima of the cycle $P$ (P is polar flux in Mx divided by 10$^{22}$), (c) the combined new precursor $\sqrt{PR}$, and (d) $P \times R$. Taken from \citet{HC19}.}
    \label{fig:prediction}
\end{figure*}

\citet{HC19} realized that the polar field $P$ at the end of the previous cycle and the decay rate $R$ at
that time can be the two precursors for predicting the next cycle, corresponding respectively to fluctuations in the
Babcock--Leighton mechanism and fluctuations in the \MC. From the data of past cycles, \citet{HC19} found
that an appropriate combination of $P$ and $R$ like $PR$ or $\sqrt{P R}$ may be a better predictor for the next cycle than
$P$ or $R$ alone. Figures~\ref{fig:prediction}(a) and (b) show the correlation of peak sunspot number of the next cycle with individual precursors $P$ and $R$ respectively. The correlations of combined precursors $\sqrt{P R}$ and $PR$ with a peak sunspot number of the next cycle are also shown in Figs.~\ref{fig:prediction}(c) and (d) respectively. As we see in Fig.~\ref{fig:prediction}, the combined precursors give a better correlation than the individual ones. In an era when there had not been a significant fluctuation in the \MC, the polar field $P$ alone may be a good enough predictor for the next cycle.  However, a combination of $P$ and $R$ may give a more complete formula for predicting the next cycle under more general circumstances.

\section{The Future: Towards bridging mean-field approaches, flux transport dynamos and full magnetohydrodynamic simulations}\label{sec7}

This review, although somewhat limited in scope due to space constraints, reinforces the view that mean-field models and the flux transport dynamo paradigm have been very useful in explaining many of the observed properties of the solar cycle, including but not limited to, the latitudinal distribution and equatorward propagation of the sunspot belt, solar cycle fluctuations, parity modulation, and have played a critical role in devising data driven models for solar cycle predictions. The mathematical structure of these solar cycle models is based on the canonical $\alpha\Omega$ dynamo equations, although in the Babcock-Leighton models, the poloidal source term is motivated from a fundamentally different perspective, or often explicitly added in an ad hoc manner to mimic the buoyant emergence of flux tubes. Moreover, these models rely significantly on a priori prescribed transport coefficients and large-scale flow profiles in stark contrast to full MHD models and magnetoconvection simulations. These appear to be orthogonal approaches and indeed, often these diverse communities have worked in silos. However, rich dividends and transformative progress may result from bridging these approaches and making use of observational constraints, when available.

The mean-field, kinematic or flux transport dynamo models rely on multiple processes. The source of the toroidal field, differential rotation, is rather well constrained by helioseismic observations \citep{howe2009}. The Babcock-Leighton poloidal source is well constrained by near surface observations and is now thought to be the dominant driver of cycle to cycle variability over at least centennial time-scales \citep{DasiEspuig10,cameron2015,bhowmik2018}; these can be adequately captured in data driven surface flux transport models or dynamo models. The meridional circulation is well observed on the solar surface and results for the solar interior are now beginning to converge as already discussed indicating a largely single cell flow threading the solar convection zone \citep{Rajaguru_Antia15,Gizon_EA_20Single_cell} in keeping with the typical profile used in flux transport dynamo models.

The origin of the meridional flow seems to be well understood in the mean-field theory and the models based on the theory agree closely with the seismologically detected single-cell circulation. The origin of the observed variability in the meridional flow is less certain however. Apart from direct modification by the Lorentz force, meridional flow of Eq.\,(\ref{modified}) is sensitive to variations in the differential rotation \citep{Rempel_05_Random_Lambda} and differential temperature \citep{Spruit_03Barocline}. The dominant mechanism for the variability remains to be identified.

Many of the important ingredients which play a crucial role in the kinematic, flux transport dynamo modelling approach are in fact well constrained and we posit this is perhaps one of the underlying reasons for its success. As already argued, such models have in fact been the first to point out the importance of single-cell meridional circulation threading the convection zone, recovering which still remains a challenge for full MHD models, although there is progress towards that direction \citep{featherstone2015}. This is just one of the examples of how the flux transport paradigm may serve as a useful guide for full MHD numerical simulations.

The reverse is also true. There is much that can be gleaned from mean field models of helical turbulent convection and full MHD simulations that are useful inputs for the flux transport models. For example, one of the  widely utilized and popular sources of the poloidal field, the mean field $\alpha$-effect cannot be directly observed in action. Although challenging and fraught with uncertainties on how to extract these transport coefficients, there are attempts to utilize full MHD models for constraining the mean-field poloidal source \citep{Kapyla06, Simard_Charbonneau_2016, Warnecke2018a, Shimada_Hotta_2022}. Another case in point is the diamagnetic pumping (already discussed in this review) and turbulent pumping of magnetic fields. Full MHD simulations point out that turbulent pumping amplitudes can be effectively comparable or faster than meridional circulation \citep{Kapyla06}. This has resulted in the construction of flux transport dynamo models that imbibe the physics of turbulent pumping of magnetic flux. Another outstanding issue is the amplitude of the effective turbulent magnetic diffusivity that is utilized in mean field or flux transport dynamo models. While this naturally arises out of turbulent convection driven by convective heat flux, its amplitude in the solar interior remains uncertain. Mixing length theory suggests strong turbulent diffusivity on the order of 10$^{12}$--10$^{13}$ cm$^2$s$^{-1}$; this sometimes introduces a problem in sustaining dynamo action, although, with low diffusion near the base of the convection zone and diamagnetic pumping, dynamo sustains \citep{Kit_Ole_12Dyn}. There are also some recent efforts \citep{KC16, Hazra2019} with high turbulent diffusivity $\sim$ 10$^{12}$ cm$^2$s$^{-1}$, which are able to produce magnetic cycles with added turbulent pumping in their model. The magnetic quenching of turbulent diffusivity in flux transport models has been demonstrated to be useful in sustaining magnetic cycles in this context too \citep{munoz2011}. This is another example where ideas from mean-field theory, magnetoconvection, and flux transport models come together to provide useful insight.

One fundamental challenge remains. Can one bring out the essence of the Babcock--Leighton mechanism -- so successfully utilized in kinematic flux transport dynamo models and now proven to drive cycle to cycle variability -- in direct numerical simulations of the solar magnetic cycle? Perhaps this is where the future lies, where ideas gleaned from all these diverse approaches may converge together.


\bmhead{Acknowledgments}
GH acknowledges IIT Kanpur Initiation Grant (IITK/PHY/2022386) and ARIEL fellowship for financial support. DN acknowledges support for the Center of Excellence in Space Sciences India at IISER Kolkata from the Ministry of Education, Government of India and multiple past students for sharing his journey of discovery. LK acknowledges financial support from the Ministry of Science and High Education of the Russian Federation. The research of ARC is supported by an Honorary Professorship from the Indian Institute of Science.

\section*{Declarations}


\begin{itemize}
\item {\bf Competing interests}
The authors declare no competing interests
\end{itemize}







\begin{appendices}





\end{appendices}


\bibliography{our_ref}


\end{document}